\title{\normalfont The finite square well: whatever is worth teaching at all is worth teaching well}
\author{
        {\em K. Razi Naqvi\/} and {\em Sigmund Waldenstr\o m\/}\\
                Department of Physics,
        Norwegian University of Science and Technology\\
        NO-7491 Trondheim, Norway\\[1ex]
         }
\date{}

\documentclass[11pt,reqno]{article}
\usepackage{amsmath,amssymb,amsfonts} 
\usepackage{amssymb}  %Some extra symbols 
\usepackage{makeidx}  %If you want to generate an index, automatically 
\usepackage{graphicx} %If you want to include postscript graphics 
\usepackage{epsfig}
\usepackage[sort&compress,numbers]{natbib}
\usepackage{pstricks}
\usepackage{soul}  
\usepackage[dvips]{curves}  
\usepackage{xcolor}
\usepackage{listings}
\usepackage{amsbsy}

\usepackage{multicol}
\usepackage{float} 
\usepackage{pstricks, pst-text}                  

\usepackage{color}
\usepackage{fancyhdr}

\newcommand{\linkcolor}{black}
\usepackage[colorlinks=true,linkcolor=\linkcolor,urlcolor=\linkcolor,citecolor=black]{hyperref}

\makeatletter
\AtBeginDocument{
\renewcommand\NAT@open{\color{black}[}}

\usepackage{url}
\usepackage[bottom]{footmisc}
 2
 0
 1
 2
 1
 2
 2
 1
 1
 2
 0
 2
 1
 1
 2
 1
 2
 2
 1
 1
 2
 1

\definecolor{hellgrau}{gray}{.8}
\definecolor{dunkelblau}{rgb}{0, 0, .7}
\definecolor{roetlich}{rgb}{1, .7, .7}
\definecolor{dunkelmagenta}{rgb}{.3, 0, .3}
\definecolor{razi01}{rgb}{.3, .7, .3}
\definecolor{razi02}{rgb}{.7, .7, .3}
\definecolor{razi03}{rgb}{.1, .5, .5}%EmeralGreen
\definecolor{razi04}{rgb}{.5, .5, .5}%DarkGrey
\definecolor{razi05}{rgb}{.6, .4, .6}%LightPurple
\definecolor{razi06}{rgb}{.1, .3, .6}%AquaMarine\hyphenation{wave-length}

\long\def\symbolfootnote[#1]#2{\begingroup%
\def\thefootnote{\fnsymbol{footnote}}\footnote[#1]{#2}\endgroup}

\graphicspath{%
    {converted_graphics/}% inserted by PCTeX
    {/}% inserted by PCTeX
    {Graphics/Chap13/}% inserted by PCTeX
    {Graphics/}% inserted by PCTeX
}
\begin{document}

\maketitle

\begin{center}
\section*{\fontfamily{phv}\selectfont\normalsize Abstract}
\label{sec:Abstract}
\end{center}
\vspace{-1ex}

A contemporary physicist would be hard put to agree entirely with the author of a 1959 textbook on quantum mechanics, who wrote: ``A second simple, one-dimensional system, somewhat divorced from reality but illustrative of the principles of the theory, is a particle in a box with finite walls." Interest in this prototypic system has not diminished over the years, and is not likely to do so in the near future, because it has now been wedded to reality and is seen as a paradigm for quantum wells and other nanosystems. The treatment of this topic in standard textbooks has changed little over the last few decades, and such changes as have come to our notice leave, in our opinion, much to be desired. The purpose of this article is to enable, even encourage, other teachers to give the finite square well the attention that it deserves; to this end, the authors provide a tutorial review that is more instructive and comprehensive than the accounts presented in numerous textbooks and dispensed in still more numerous online resources, but does not make any additional demands on the mathematical abilities of the student. Given the elemental nature of the topic, the authors claim no originality; with nothing more than the winnowing fork in their hands, they can only clear the threshing floor, gather the wheat into the barn, and let others decide whether or no the chaff should be burnt with unquenchable fire. \\[6ex]

%His winnowing fork is in his hand, and he will clear his threshing floor, gathering his wheat into the barn and burning up the chaff with unquenchable fire." Matthew 3:12

{\large
\begin{multicols}{2}

\section{Introduction}

When teaching the finite square well to students, teachers would do well to follow Lord Chesterfield's advice to his son \cite[p.~41]{Chesterfield1893Letters1}: ``In truth, whatever is worth doing at all is worth doing well, and nothing can be done well without attention: I therefore carry the necessity of attention down to the lowest things, even to \ldots ". We have omitted the last three words of his remark because they should be replaced, in the present context, by the phrase ``the last significant digit". One of our aims is to reiterate what was pointed out (in the 1970s) after the advent of hand-held electronic calculators \cite{PhillipsMurphyAJP,MurphyPhillips1976AJP,Memory1977AJP}: finding the eigenenergies of a finite square well provides a springboard for introducing a physics student to equations whose solutions can only be found through an iterative or recursive technique. To illustrate the operations involved in such calculations, we use a commercial spreadsheet (Microsoft$^{\circledR}$ Excel$^{\circledR}$). The widespread availability of spreadsheet software has made it possible to go beyond the graphical approaches which were developed in the days when computing facilities were primitive and scarce. We also show, by considering a specific case (a square well whose uppermost energy level is close to the top of the well), that occasionally the desired result may be obtained more easily if one is prepared to supplement the resources of Excel with some cerebration.

We conclude the preamble by citing the source of the remark quoted in the Abstract \cite[p.~41]{Sherwin1959QM} and by recalling
a recent article where the utilitarian relevance of the problem is acknowledged \cite{BarsanPhilMag2014}.

\section{Notation and Basic Relations}

Our purpose will be best served by adhering to the notation used in a book on modern physics which has gone through three editions and several reprints; the symbols $U$ and $L$ stand in this book for the height and width of the well, respectively; $m$ and $E$ $(0<E<U)$ denote the mass and energy of the particle, respectively, and the subsequent treatment requires the introduction of two more symbols \cite[pp.~209--212]{Serway2005Modern}:
\begin{align}
k&=\left (2mE/\hbar^2 \right )^{1/2},\label{eq:kay}\\
\alpha &=\left [2m(U-E)/\hbar^2 \right ]^{1/2}.\label{eq:alpha}
\end{align}

The allowed energy levels $\bigl (E_n^{[\infty]}\bigr )$ of a particle confined to an {\em infinite\/} square well (of width $L$) are given by the expression 
\begin{equation}\label{eq:ISW1}
E_n^{[\infty]}=n^2\frac{\hbar^2 \pi^2}{2mL^2}= n^2E_1^{[\infty]},\mbox{ say},
\end{equation}
where $E_1^{[\infty]}$ is the ground state energy and $n$ denotes the quantum number $(n=1,2,\ldots)$. For the {\em finite\/} square well, the wavefunction penetrates the classically forbidden region, and the so-called penetration depth (or decay length) is defined as
\begin{equation}\label{eq:delta1}
\delta = \frac{1}{\alpha}=\frac{\hbar}{\left [2m(U-E)\right ]^{1/2}}.
\end{equation}
Solution of the time-independent Schr\"{o}dinger equation for the finite well leads to the conclusion that allowed values of $E$ are those which satisfy the following transcendental equation \cite[pp. 211--2]{Serway2005Modern}:
\begin{equation}\label{eq:trans1}
\frac{(\alpha/k)\cos(kL)-\sin(kL)}{(\alpha/k)\sin(kL)+\cos(kL)}=-\frac{\alpha}{k}.
\end{equation}
One sees from Eqs.~\ref{eq:kay} and  ~\ref{eq:alpha} that both $\alpha$ and $k$ depend on the value of $E$. It is convenient to denote the ratio $\alpha/k$ by $\Gamma$, to note that 
\begin{equation}\label{eq:DefGamma}
\Gamma =\frac{\alpha}{k}=\sqrt{\frac{U}{E}-1},
\end{equation}
and go on to rearrange Eq.~\ref{eq:trans1} as
\begin{equation}\label{eq:trans2}
\tan(kL) =\frac{2\Gamma}{1-\Gamma^2},
\end{equation}
so that one may avail oneself of the trigonometric identity
\begin{equation}
\tan A =\frac{2\tan (A/2)}{1-\tan^2(A/2)},
\end{equation}
and express Eq.~\ref{eq:trans2} as
\begin{equation}\label{eq:trans3}
\frac{2\tan \xi}{1-\tan^2\xi} = \frac{2\Gamma}{1-\Gamma^2},
\end{equation}
where 
\begin{equation}\label{eq:omega1}
\xi = \frac{kL}{2}=\frac{L}{2\hbar}\sqrt{2mE}.
\end{equation}

Now, Eq.~\ref{eq:trans3} is simply a second-order equation in $\tan\xi$ with solutions of even and odd parity:
\begin{subequations}
\begin{align}
\tan\xi &=\Gamma \label{eq:graphic1}\\
\tan\xi &=-1/\Gamma \label{eq:graphic2}
\end{align}
\end{subequations}
The two solution cannot coincide, for that would imply $\tan^2\xi=-1$. The energy is now given by the relation
\begin{equation}\label{eq:energy1}
E_n= \frac{2\hbar^2}{mL^2}\,\xi_n^2 = E_1^{[\infty]}\left (\frac{2}{\pi}\right )^2\,\xi_n^2.
\end{equation}

Multiplication of both sides of Eq.~\ref{eq:trans1} with the denominator on the left-hand side, transposition of the right-hand side of the resulting equation to the opposite side, and multiplication by $k^2$ leads to the result shown below:
\begin{equation}\label{eq:Razi1}
(\alpha^2-k^2)\sin (kL) + 2\alpha k \cos (kL) =0.
\end{equation}

Let us introduce the symbol 
\begin{equation}\label{eq:DefR}
R\equiv \surd(k^2+\alpha^2)=\frac{\surd(2mU)}{\hbar},
\end{equation}
and use it for defining $\eta$ through the relations
%\begin{equation}\label{eq:Razi2}
%\frac{\alpha^2-k^2}{R^2}\sin (kL) + \frac{2\alpha k}{R^2} \cos (kL) =0.
%\end{equation}
\begin{subequations}
\begin{align}%\label{eq:gamma1}
\sin \eta &= \frac{k}{R}=\frac{k\hbar}{\surd{(2mU)}},\\
\noalign{\noindent\mbox{and}}
\cos \eta &= \frac{\alpha}{R}=\frac{\alpha\hbar}{\surd{(2mU)}}.
\end{align}
\end{subequations}
When one recalls the identities
\begin{subequations}
\begin{align}
\sin(2\eta)&=2\sin\eta\cos\eta=\frac{2\alpha k}{R^2},\\
\cos(2\eta)&=\cos^2\eta-\sin^2\eta=\frac{\alpha^2-k^2}{R^2},
\end{align}
\end{subequations}
one sees without further ado that Eq.~\ref{eq:Razi1}, after it is divided by $R^2$, can be recast as
\begin{equation}\label{eq:kL2eta}
\sin(kL+2\eta)=0,
\end{equation}
or, after replacing $kL$ by $2\xi$, as
\begin{equation}\label{eq:Razi2}
\sin [2(\xi + \eta)] =0.
\end{equation}
%whose graphical representation requires nothing more than plotting a sinusoid.

For reasons that will become transparent shortly, we observe that Eq.~\ref{eq:Razi2} implies that
\begin{equation}\label{eq:Razi3}
2\xi  = (n+ 1)\pi - 2 \eta
\end{equation}
where $n = 0, 1, 2, \cdots$. Replacing $2\xi$ by $kL$ and $\eta$ by $\sin^{-1}[k\hbar/\surd{(2mU)}]$, we arrive at the relation
\begin{equation}\label{eq:Razi4}
kL = (n+1)\pi-2\sin^{-1}[k\hbar/\surd{(2mU)}],
\end{equation}
which has been available to the English-reading public since the publication (in 1958) of the first edition of the translated version of the splendid textbook on quantum mechanics by Landau and Lifshitz \cite{LandauLifshitz1965QM}. It is most convenient for our purpose to refer to the abridged version of their book \cite{LandauLifshitz1974SQM}, and we point out that, if our symbols $L$ and $U$ are replaced by $a$ and $U_0$, our Eq.~\ref{eq:Razi4} becomes identical with their Eq.~1 on p.~80.

We will continue to use our own notation for describing the approach followed by Landau and Lifshitz (L\&L, for short). They stated that their Eq.~1 can be recast in a more convenient form if one introduces two new variables, one being $\xi$ and the other
\begin{equation}\label{eq:Landau1}
\gamma=(\hbar/L)\surd{(2/mU)}.
\end{equation}
When $n$ is even, the resulting equation is 
\begin{equation}\label{eq:Landau2}
\cos\xi =\pm \gamma \xi,
\end{equation}
{\em and the roots for which $\tan\xi>0$ must be taken\/}. When $n$ is odd, we have
\begin{equation}\label{eq:Landau3}
\sin\xi =\pm \gamma \xi,
\end{equation}
{\em and the roots for which $\tan\xi<0$ must be taken\/}. 

%Before commenting further on the approach adopted by L\&L, we pause to introduce 

We will now introduce a parameter (called the {\em power\/} or {\em strength\/} of the well) that has been used by many authors \cite{Pitkanen1955AJP,Krivchenkov1960Problems,Garrett1979AJP,Barker1991AJP,Sprung1992EJP,Sprung1996AJP,Reed1990AJP}; we will denote it by the symbol $P$, and it is enough to point out that $P\equiv 1/\gamma$, and to note, for later use, that
\begin{equation}\label{eq:Power1}
\Gamma=\left [\frac{P^2}{\xi^2}-1 \right ]^{\frac{1}{2}},
\qquad (0\leq \xi <P ).
\end{equation}
Finally, putting $E=0$ in Eq.~\ref{eq:delta1}, and denoting the resulting expression by $\Delta$, we write
\begin{align}
\Delta &=\frac{\hbar}{\surd (2mU)},  \label{eq:delta2}\\
\noalign{\noindent\mbox{and note that}}
\frac{\Delta}{L}&=\frac{\gamma}{2}.            \label{eq:delta3}
\end{align}

\section{A Retrospect}

The first edition (in Russian) of L\&L's book on non-relativistic quantum mechanics was published in 1947; its English translation appeared in 1958 \cite{LandauLifshitz1965QM}. Gol'dman, Krivchenkov, Kogan and Galitskii (hereafter GKKG) used L\&L's approach but added a graphical solution in their well-known book, whose  English translation was published in 1960 \cite{Krivchenkov1960Problems}. Among the authors who wrote directly in English, Pitkanen \cite{Pitkanen1955AJP} appears to have been the first to propose some graphical representations based on Eqs.~\ref{eq:Landau2} and \ref{eq:Landau3}, and one of these representations is equivalent to that displayed in Fig.~20 of Ref.~\cite[p.~60]{Krivchenkov1960Problems}; it is not inconceivable that, en route, he also deduced our Eq.~\ref{eq:Razi4}, but found it unnecessary to bring it to the readers' attention because in those days teachers, more occupied with finding user-friendly graphical procedures, spent a great deal of ingenuity on devising plots which used simple continuous curves (such as sinusoids) and straight lines. In an age when a contemporary student can plot, with a few keystrokes and a couple of mouse clicks,  half a dozen curves, each with a different colour, the relevant question to ask is not which graph is easier to generate but which is more instructive than other claimants to a student's attention.

Cantrell \cite{Cantrell1971AJP}, who referred to Landau and Lifshitz \cite{LandauLifshitz1965QM}, concluded his contribution with the following comment: 

\vskip 8 truept
\leftskip  = 10 pt
{\small
The only pitfall of the method is in remembering which diagram (sine or cosine) is appropriate for a given situation, and where the ``forbidden" regions are on the diagram. While a complete mnemonic would be too lengthy to be useful, it is possible to give a simple rule; the {\em sine\/}, which is an {\em odd\/} function, is associated with the {\em odd\/}-parity states in the symmetrical well \ldots, and with {\em negative\/} values of $\tan ka$ [$\tan (kL)$, in our notation], for the purposes of this construction. 
}

\vskip 8 truept
\leftskip  = 0 pt

Now, {\em spoon-feeding\/} is usually defined as `providing someone with so much help or information that they do not need to think for themselves' \cite{Pearsall1998New}. Students who are unable to figure out, after thinking hard (and, if necessary, long) the restrictions on the sign of $\tan \xi$ should be encouraged to change to a field of study better suited to their particular talents. A key reference in our bibliography \cite{Barker1991AJP} has a space filler in which Sylvia Plath is quoted as saying, ``The day I went into physics class was death"; one only hopes that the depression which tormented the prodigiously talented Plath has not been traced by a biographer to tiresome mnemonics composed by a physics teacher desperate to narrow the gulf between ``The Two Cultures".

An equation of the same form as Eq.~\ref{eq:Razi1} was derived by Reed \cite{Reed1990AJP}, and subsequently shown by Sprung, Wu and Martorell (SW\&M) to be equivalent to the compact form displayed in Eq.~\ref{eq:Razi2}; in their first publication on the topic \cite{Sprung1992EJP}, SW\&M claimed (justifiably) to have derived no more than a ``new and very accurate analytic solution \ldots for the energy levels of the finite square well potential, {\em in the form of a rapidly convergent series in inverse powers of the strength\/}" (emphasis added). The discovery of Eq.~\ref{eq:Razi2} was attributed by Aronstein and Stroud \cite{AronsteinPRA2000} to SW\&M \cite{Sprung1992EJP}, who began the sequel to their first paper with a claim for the derivation of the simple formula itself \cite{Sprung1996AJP}:

\vskip 8 truept
\leftskip  = 10 pt
{\small
Some time ago we provided a simple analytical formula for the bound state energies of the finite square well potential. We also showed how our solution is related to the graphical construction of Pitkanen. We believe that these methods are superior to those presented in most textbooks, and should help in presenting this canonical example in introductory quantum mechanics classes. 
}

\vskip 8 truept
\leftskip  = 0 pt

We conclude this section by pointing out that explicit analytical expressions for the energies of the finite square well are also available \cite{Siewert1978JMP,PaulNkemzi2000JMP,Aronstein2000JMP,Bluemel2005JPA}, but will not be considered here, because their derivation and further manipulation requires techniques not found in the mathematical repertoire of a physics undergraduate; we will, however, quote and use an approximation that is valid for the case of small $\gamma$ (see below).

\section{Calculation of Eigenenergies}

Since we wish to comment on the works of other authors, we will use two conventions for actual calculations, and measure energies in eV and distances in nm, or work with dimensionless quantities; in the latter case, energies will be measured in terms of $\varepsilon\equiv \hbar^2/mL^2$ and lengths in terms of $L$.

\subsection{First Choice: Eq.~7}

If one's sole aim is to find the eigenenergies, one need go no further than Eq.~\ref{eq:trans2}, since it is a single equation in a single unknown $E$; in a graphical representation, one would look for the intersections of the curve $y=\tan (kL)$, which shows multiple discontinuities, with the curve $y=2\Gamma/(1-\Gamma^2)$, which has only one singularity (at $E=U/2$).

Let us rewrite Eq.~\ref{eq:trans2} in terms of the symbols representing the properties of the particle and the well, 
\begin{equation}\label{eq:Lindberg1}
\frac{2\sqrt{(U-E)E}}{2E-U}=\tan\left ( \sqrt{\frac{2mEL^2}{\hbar^2}}\right ),
\end{equation}
and consider a particular case that was analyzed recently by Lindberg \cite{LindbergFSW}: a well of height $U=25 \mbox{ eV}$ and $L=0.5 \mbox{ nm}$. Lindberg prepared a three-column table and put values of energy $E$ in steps of 0.1 eV from 0.1 to 25 eV (in the first column) and the corresponding values of the two sides of Eq.~(\ref{eq:Lindberg1}) in the other two columns. He plotted the values of the second and third columns against $E$, looked for intersections of the two curves, and adjusted the energy at each intersection ``to get something close to a perfect fit, i.e. LHS = RHS".

The procedure followed by Lindberg is simple and robust, and adequate if the teacher's aim is to obtain results correct to a few decimal places. A minor blemish is that the plot, which exhibits five crossings, does not separate levels belonging to different parities, since the first three crossings occur when the left-hand side is negative ($0<E<U/2$), and the last two when the left-hand side is positive (for $U/2<E<U$).

We believe that the pedagogical value of the exercise can be substantially enhanced by informing the students that other, more efficient ways of finding precise values of the roots are available. We will describe these methods below before considering the advantages of using other equations for finding the eigenenergies.

\subsection{Some Root-Finding Procedures}

The {\sf Analysis ToolPak} of Excel contains two tools, namely {\sf Goal Seek} and {\sf Solver} \cite{GoalSeekSolver}, which can find (in most cases) the roots of the equation $f(x)=g(x)$ within a stated precision; the user must supply an initial guess for the root, say $x_0$, calculate the value of the difference $\delta (x_0)=f(x_0)-g(x_0)$, and let the tool search for the value of the argument for which $\delta$ attains a value sufficiently close to zero (that is, less than a value set by the user). {\sf Goal Seek} and {\sf Solver} are superb tools indeed, but it seems advisable to inform a novice, who is likely to view each of these as a magic wand, that they are not foolproof (see below) and other methods, not making explicit use of {\sf Analysis ToolPak}, are available. 

In the mid-1970s, when personal computers were not yet commonplace, the availability of hand-held electronic calculators (also called electronic slide rules) was seen by some teachers as an opportunity to introduce their students to root-finding procedures. We will refer to three brief and lucid contributions \cite{PhillipsMurphyAJP,MurphyPhillips1976AJP,Memory1977AJP}, which focussed primarily on the finite square well, and will take it for granted that those who need some instruction would consult these sources, and other references cited therein.  Among the many methods that could be brought to bear on our problem, only two will be considered here, namely iteration and bisection.

The basic idea behind the bisection scheme has been explained in the context of the finite square well in a recent book on computational physics \cite[pp.~134--135]{Landau2011Survey}; the Internet is a rich source of information as to how this method may be implemented by using Excel, and it will be enough to cite one resource here \cite{Bisection2011Colorado}, which has tested by the authors and found to be satisfactory.

As it stands, Eq.~\ref{eq:trans2} is not of the form $x=f(x)$, and thus not well-suited to an iterative treatment, but it can be easily transformed, as shown above, in two different ways into a pair of equations with the desired form; details concerning an iterative extraction of the roots of the transformed equations are given below.

The five allowed energy levels of the square well under consideration are displayed in Table~1. The first four roots of Eq.~\ref{eq:Lindberg1} found by using bisection, {\sf Goal Seek} and {\sf Solver} concurred with one another. It is crucial to note that {\sf Solver} failed to find $E_5$, regardless of how close the initial guess (in eV), to be denoted as $E_k^{\ast}$, was to the correct value, but {\sf Goal Seek} and bisection returned the correct answer, so long as the input from the user was chosen judiciously; as to what {\em judicious choice\/} means in this context, there is no substitute for personal experience, but a few remarks may not be amiss here.
Though we will speak of only {\sf Goal Seek}, the following remarks are equally applicable to {\sf Solver}.

\begin{table}[H]
\caption{Energy levels of a finite square well ($U=25$ eV, $L=0.5$ nm).  The calculations used NIST data \cite{Constants}, which led to the following values: $\hbar c=1.97327\times 10^2 \mbox{ eV$\cdot$nm}$ and $mc^2=5.10999\times 10^5 \mbox{ eV}$.}
%\hline
\vspace{1ex}
\centering
\begin{tabular}{lcr}
%\toprule
%\cmidrule(r){1-2}
\hline\hline\\[-1ex]
{\large$n$}&\hspace{6ex} & {\large$E_n/\mbox{eV}$}  \\ [0.5ex]
%\midrule
\hline
 1 & &1.12294  \\ 
 2 & &4.46186  \\  
 3 & &9.90751  \\
 4 & &17.16578  \\
 5 & &24.78411  \\
\hline\hline
%\bottomrule
\end{tabular}
\end{table}
\label{table:Lindberg}

%http://www.math.ubc.ca/~israel/m210/lesson7.pdf

Let us consider the third eigenvalue, and note that if one takes $E_k^{\ast}=9.3$, which is much closer to $E_3$ than to $E_2$, {\sf Goal Seek} converges to the value of $E_2$; furthermore, for many choices of  $E_k^{\ast}$, for example 9.4 or 12.4, {\sf Goal Seek} fails to find any of the eignevalues. An alert student should be able to explain these observations (and others of a similar nature) by examining a graph in which $y=2\sqrt{(U-E)E}/(2E-U)$ and $y=\tan\surd{(2mEL^2/\hbar^2)}$ are plotted against $E$.

\subsection{Separating States of Even and Odd Parity}

As already stated, to anyone who has access to modern computing facilities, the difference between plotting two curves or three is not an important consideration. It is worthwhile, therefore, to employ a different graphical procedure for illustrating the calculation. Figure~\ref{fig:FigureOne} is a plot based on Eqs.~\ref{eq:graphic1} and \ref{eq:graphic2}; $y=\tan \xi$, $y=\Gamma$ and $y=-1/\Gamma$ have been plotted against $\xi$. One sees from the figure that three solutions of even parity and two of odd parity are permitted, and that the even/odd parity solutions correspond to the intersections in the upper/lower half of the plot. 

\end{multicols}

\begin{figure}[t!] % float placement: (h)ere, page (t)op, page (b)ottom, other (p)age
  \centering
  % file name: H:/Sigmund FSW/FigureOne.eps
  \includegraphics[bb=0 0 341 238,width=12cm,height=8.37cm,keepaspectratio]{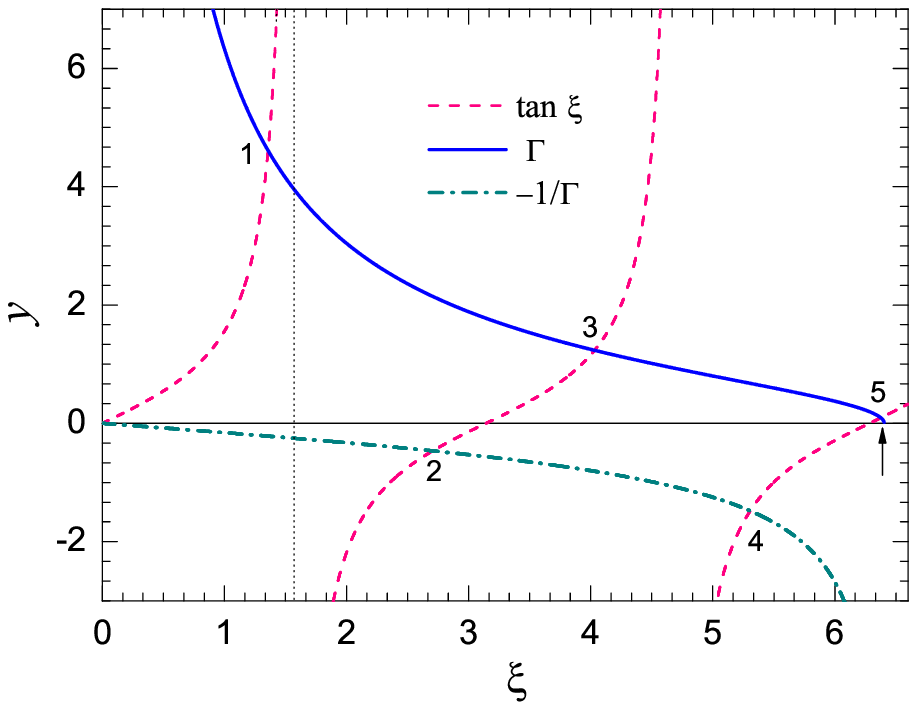}
\vspace*{-2ex}
  \caption{Graphical representation of  Eqs.~\ref{eq:graphic1} and \ref{eq:graphic2} ($U = 25 \mbox{ eV}$, $L = 0.5 \mbox{ nm}$). The intersections of $y=\tan\xi$ and $y=\Gamma$, labelled as 1, 3, and 5, belong to even parity states; the intersections of $y=\tan\xi$ and $y=-1/\Gamma$, labelled as 2 and 4, pertain to odd parity states. The vertical arrow indicates the abscissa ($\xi_{\rm max}\equiv P=6.4039\cdots $) at which $y=\Gamma$ crosses the horizontal axis. The dotted vertical line is the asymptote to $y=\tan \xi$ at $\xi=\pi/2$.}
  \label{fig:FigureOne}
\end{figure}

\begin{multicols}{2}

The pivotal role played by the parameter $\gamma$ (or its reciprocal $P$) becomes apparent when one considers two limiting cases: $P\to \infty$ and $P\to 0$. A student ought to be able to deduce, merely by examining the locations of the intersections, that in the former case, one recovers Eq.~\ref{eq:ISW1}, and that if $P<\pi/2$, only one energy level of even parity will be allowed. We suggest that the task of supplying the missing steps in some remarks made by L\&L and GKKG may also be used as student exercises. In treating the case $\gamma \ll 1$, L\&L argue as follows: ``In particular, for a shallow well in which $U_0 \gg \hbar^2/ma^2$, we have $\gamma \ll 1$ and equation (3) [our Eq.~\ref{eq:Landau3}] has no root. Equation (2) [our Eq.~\ref{eq:Landau2}] has one root (with the upper sign on the right-hand side), $\xi\cong 1/\gamma - 1/2\gamma^3$. Thus the well contains only one energy level, \ldots , which is near the top of the well." GKKG also made this remark (without an explanation), and went on to state that the number of levels for arbitrary values of $U$ and $L$ will be equal to $N$, where $N$ follows from the relation
$$
N >\frac{2P}{\pi} > N-1.
$$
In modern notation the above result is stated as
\begin{equation}\label{eq:floor1}
N=1+\left \lfloor \frac{P}{\pi/2}\right \rfloor,
\end{equation}
where $\lfloor x \rfloor$ is the floor function defined here as the largest integer smaller than $x$.

We turn next to the problem of finding the abscissa for each intersection, and will consider four options: {\sf Goal Seek}, {\sf Solver}, bisection and iteration. Phillips and Murphy have discussed the solution of the equation $x=\tan(\beta x)$ by iteration \cite{PhillipsMurphyAJP}, and their discussion can be easily adapted to Eqs.~\ref{eq:graphic1} and \ref{eq:graphic2}. It will be enough to state here that $E_1$ and $E_3$ were determined by transforming Eq.~\ref{eq:graphic1} as $\xi=\tan^{-1}(\Gamma)$, and Eq.~\ref{eq:graphic2} was rephrased as $\xi=\tan^{-1}(-1/\Gamma)$ for determining $E_2$ and $E_4$; for the uppermost level, $\Gamma$ was considered to be the independent variable, and Eq.~\ref{eq:graphic1} treated as $\Gamma = f(\Gamma)$. All four methods gave results which were concordant, except that {\sf Solver} failed to find the value of $E_5$ (which is exceptional in the sense that it lies close to the top of the well).

The analysis presented above was developed by the authors as an alternative to the use of Eqs.~\ref{eq:Landau2} and \ref{eq:Landau3}, but other teachers may prefer to use the latter pair. Appendix~I deals with the application of the iterative approach to these equations.

\section{Finite and Infinite Wells}

Since the wavefunction of a particle subject to a finite square well potential extends beyond the boundaries \cite[pp.~209--212]{Serway2005Modern}, decaying exponentially at a rate $\delta$, Garrett \cite{Garrett1979AJP} proposed that the approximate energy eigenvalues (to be denoted by ${\cal E}_n$) of the finite square well can be approximated by adapting Eq.~\ref{eq:ISW1}, valid for the energies of an infinite square well, as follows: 
\begin{equation}\label{eq:GarrettE}
{\cal E}_n = \frac{n^2\pi^2\hbar^2}{2m(L+2\delta_n)^2}.
\end{equation}
Since $\delta_n$ is an energy-dependent quantity, Eq.~\ref{eq:GarrettE} is, in fact, an implicit relation for ${\cal E}_n$ that must be solved numerically for a given value of $n$ by using the relations
\begin{align}
\delta_n^{(k)} &= \frac{1}{\alpha}=\frac{\hbar}{\sqrt{2m\left [U-{\cal E}_n^{(k-1)}\right ]}},\label{eq:delta-n}\\       
\noalign{\noindent\mbox{with $k=1,2,\ldots$, and}}
{\cal E}_n^{(k)} &= \frac{n^2\pi^2\hbar^2}{2m\left[L+2\delta_n^{(k)}\right]^2},\label{eq:GarrettE2}
\end{align}
where ${\cal E}_n^{(k)}$ and $\delta_n^{(k)}$ each denote the $k$-th iterate. All that is needed now is an appropriate guess for ${\cal E}_n^{(0)}$, and Garrett suggested that one should take ${\cal E}_n^{(0)}=E_n^{[\infty]}$.

The authors of Ref.~\cite{Serway2005Modern}, hereafter referred to as SM\&M, illustrated Garrett's approach by calculating ${\cal E}_1^{(k)}$ for a well with $U=100 \mbox{ eV}$ and $L=0.2 \mbox{ nm}$.  To start the iteration, they set ${\cal E}_1^{(0)}=0$  (not ${\cal E}_1^{(0)}=E_1^{[\infty]}$, as proposed by Garrett!), and obtained thereby
\begin{equation}\label{eq:delta-0}
\delta_1^{(1)} = \frac{\hbar}{\sqrt{2mU}} = 0.0195 \mbox{ nm},
\end{equation}
and
\begin{equation}\label{eq:EnerIt1}
{\cal E}_1^{(1)} = \frac{\pi^2\hbar^2}{2m\left [L+2\delta_1^{(1)}\right]^2} = 6.58 \mbox{ eV}.
\end{equation}
They went on to calculate 
\begin{equation}\label{eq:delta-2}
\delta_1^{(2)} = \frac{\hbar}{\sqrt{2m\left [U-{\cal E}_1^{(1)}\right]^2}} = 0.0202 \mbox{ nm},
\end{equation}
inserted it into Eq.~\ref{eq:GarrettE2} to find ${\cal E}_1^{(2)}=6.53 \mbox{ eV}$, and stated: ``The iterative process is repeated until the desired accuracy is achieved. Another iteration gives the same result to the accuracy reported. This is in excellent agreement with the exact value, about 6.52 eV for this case." 

The above calculation and the ensuing conclusions are so farcical as to be fit only for a homework exercise designed to test a student's ability for plugging numerical values into a formula and calculating the answer.  A student who is given this task would be able, if not incapacitated by ``tumid apathy and no concentration'',\symbolfootnote[4]{These words are form T. S. Elliot's {\em Burnt Norton.\/}} to reproduce the values of $\delta_1^{(1)}$, ${\cal E}_1^{(1)}$ and $\delta_1^{(2)}$ given above (Eqs.~\ref{eq:delta-0}--\ref{eq:delta-2}), and would probably mutter ``So far so good"; if the student has set up the spreadsheet formulas correctly, (s)he would find that ${\cal E}_1^{(2)} =  6.5071 \mbox{ eV}$, and ${\cal E}_1^{(3)} =  6.5080 \mbox{ eV}$. The discrepancy between these values and the results of SM\&M (according to whom ${\cal E}_1^{(2)}=6.53 \mbox{ eV} = {\cal E}_1^{(3)}$, and $E_1=6.52 \mbox{ eV}$) might prompt a student (or a teacher) to calculate $E_1$ for herself/himself, and such a person would find that $E_1=6.557 \mbox{ eV}$. If one repeats the calculation and takes ${\cal E}_1^{(0)}=E_1^{[\infty]}$, one would find ${\cal E}_1^{(2)} =  6.508 \mbox{ eV}$, which shows that convergence is rather rapid in this particular case, and that, whatever the initial guess, one ends up with a result that is close to 6.508 eV. The results of our own calculation have been reproduced in Figure~\ref{fig:Serway}.

\end{multicols}

\begin{figure}[ht]
\begin{minipage}[b]{0.5\linewidth}
\centering
\includegraphics[width=\textwidth]{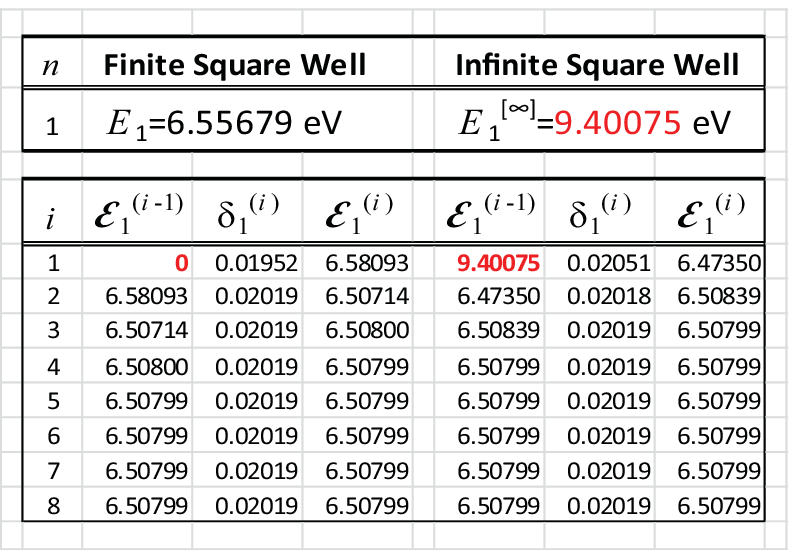}
\caption{Excel output showing the \mbox{application} of Garrett's approximation to a square well with $U=100 \mbox{ eV}$ and  $L=0.2 \mbox{ nm}$. The first  three columns show the results obtained by using ${\cal E}_1^{(0)}=0$; the last three, by using ${\cal E}_1^{(0)}=E_1^{[\infty]}$. }
\label{fig:Serway}
\end{minipage}
\hspace{0.15cm}
\begin{minipage}[b]{0.5\linewidth}
\centering
\includegraphics[width=\textwidth]{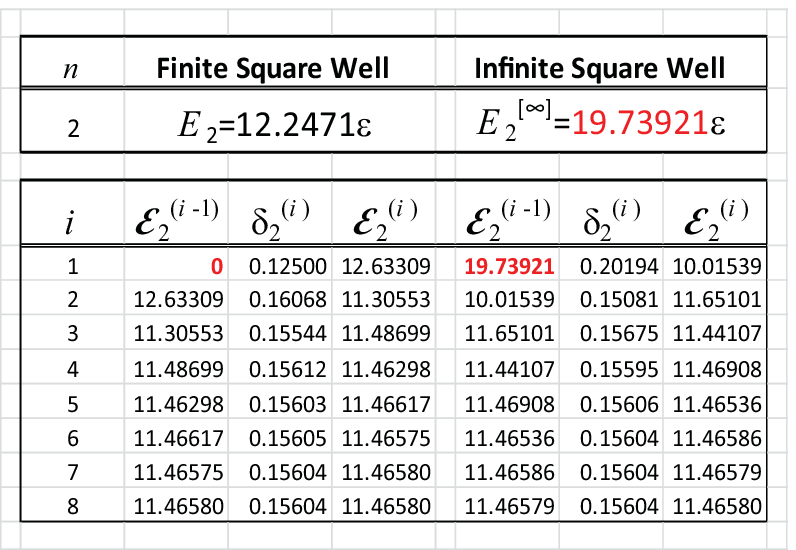}
\caption{Excel output showing the application of Garrett's approximation to a square well with $U=32\varepsilon$, where $\varepsilon = \hbar^2/(mL^2)$. The first  three columns show the results obtained by \mbox{using} ${\cal E}_2^{(0)}=0$; the last three, by \mbox{using} ${\cal E}_2^{(0)}=E_2^{[\infty]}$.}
\label{fig:Garrett}
\end{minipage}
\end{figure}
\vspace*{-10ex}

\vspace*{10ex}
\begin{multicols}{2}

Garrett (who used the symbol $a$ for our $L$) considered a well of depth $32\,\varepsilon$, put ${\cal E}_1^{(0)}=E_1^{[\infty]}$, and he too stopped at the second iteration, not only for calculating the energy of the first level, where convergence is indeed rapid, but also for calculating ${\cal E}_2$. He found the exact energy to be $E_2=12.0\,\varepsilon$, took ${\cal E}_2^{(0)}=E_2^{[\infty]}=2\pi^2\varepsilon$, and arrived at the result ${\cal E}_2^{(2)}=11.65\,\varepsilon$. The good agreement between ${\cal E}_2^{(2)}$ and $E_2$ is illusory. One can verify, in the first place, that the exact value reported in Ref.~\cite{Barker1991AJP}, $E_2=12.2471\,\varepsilon$, is correct, and (secondly) that, if the iteration is continued, convergence is reached after seven steps (see Figure~\ref{fig:Garrett}) and yields ${\cal E}_2^{(m)}=11.4658\,\varepsilon$ ($m\geq 7$). It is curious that the authors of Ref.~\cite{Barker1991AJP}, who were interested in comparing the relative error in their first-order approximation with that in Garrett's approximation, did not press the latter approximation to its logical end.

The well considered by Garrett has only three bound states. He stated: ``Clearly, this method cannot be used for the $n=3$ state since $E_0>V_0$ [$E_3^{[\infty]}>U$, in our notation]. \ldots The value of ${\cal E}_3^{(2)}$ in [his] Table~1 was obtained by taking a linear extrapolation of $\delta_1^{\prime}$ and $\delta_2^{\prime}$ to get $\delta_3=0.171a$". Garrett's symbols $\delta_1^{\prime}$ and $\delta_2^{\prime}$ correspond to our $\delta_1^{(1)}$ and $\delta_2^{(1)}$, respectively, and his $\delta_3$, which corresponds to our $\delta_3^{(1)}$, has no superscript because he did not continue the iteration. This stratagem led him to the result ${\cal E}_3^{(1)}= 24.7\,\varepsilon$, in excellent agreement with the exact value $E_3=25.9 \mbox{ eV}$. If Garrett had proceeded to the next step of iteration, using ${\cal E}_3^{(1)}$ as the input, he would have found ${\cal E}_3^{(2)}= 19.17\,\varepsilon$! Since the iteration under consideration always converges to the same result for {\em any\/} choice of ${\cal E}_n^{(0)}$ that is lower than U (see Figure~\ref{fig:Garrett}), the only legitimate result is that found {\em after convergence is reached\/}; one must conclude that Garrett's approximation does not provide the accuracy he claimed.

Although Garrett's {\em scheme\/} does not provide useful results, the {\em idea\/} of replacing a finite square well by a wider infinite well 
can be placed on a sound basis. The reader will have noticed that Garrett's own proposal amounts to replacing a finite square well ($N$ allowed energy levels) with $N$ infinite wells, each of a different width. An obvious amendment is to replace the finite well with a single infinite well of a {\em unique\/} width $\widetilde{L}$; to this end, we remind ourselves of Eqs.~\ref{eq:delta1}, \ref{eq:delta2} and \ref{eq:delta3}, and  define
\begin{align}
\widetilde{L}   & = L+2\Delta = L(1+\gamma), \label{eq:width1}\\
\noalign{\noindent\mbox{and}}
\widetilde{E}_n & =\frac{n^2\pi^2\hbar^2}{2m\widetilde{L}^2}.\label{eq:width2}
\end{align}
One expects that the approximation would work well when $E_n\ll U$, and that it would gradually worsen as $n$ becomes larger, and this is indeed found to be the case  \cite{Barker1991AJP}.

Barker and co-authors \cite{Barker1991AJP} noted that a Taylor expansion of Eq.~\ref{eq:Landau2} about $n\pi/2$, 
\begin{align}
&\cos\left (\xi-\frac{n\pi}{2} \right )= -\frac{\sin(n\pi/2)}{1!} \left (\xi-\frac{n\pi}{2} \right )+\nonumber\\
&\quad\quad\frac{\sin(n\pi/2)}{3!}
\left (\xi-\frac{n\pi}{2} \right )^3 + \cdots
\end{align}
where $n$ is an odd integer, could be used to develop approximations for Eq.~\ref{eq:Landau2}, and likewise for Eq.~\ref{eq:Landau3}.
If one retains only the first-order term, one arrives at the result
\begin{equation}\label{eq:BarkerBox}
\xi_n= (n\pi/2)[P/(P+1)],
\end{equation}
which leads, by virtue of Eq.~\ref{eq:energy1}, to Eq.~\ref{eq:width2}.

Paul and Nkemzi \cite{PaulNkemzi2000JMP} showed that the explicit expression for $E_n$ can be expanded as a series in powers of $\gamma$; when their result is corrected for an algebraic error, identified by Aronstein and Stroud \cite{Aronstein2000JMP}, the expression for the energy comes out to be
\begin{equation}\label{eq:PaulNkemziFirst}
E_n(\gamma) = \frac{n^2\hbar^2 \pi^2}{2m\widetilde{L}^2} + {\cal O}(\gamma^3).
\end{equation}

An altogether different motivation (the validity of Ehrenfest's theorem) led Rokhsar \cite{Rokhsar1996AJP} to the relation $\widetilde{L}= L(1+\gamma)$.

The results displayed in Figs.~\ref{fig:Serway} and \ref{fig:Garrett} do not corroborate the conclusion, reached in a recent publication \cite{BarsanDragomir2012}, that Garrett's iteration, if carried {\em ad infinitum\/}, leads to the result stated in Eq.~\ref{eq:width2}. This can only happen if $\delta_k^{(\infty)}$ becomes independent of $k$ and equals $\Delta$, which requires fulfillment of the absurd condition ${\cal E}_k^{(\infty)}=0$ .

\section{Concluding Remarks}

The reader who is totally satisfied with the sterling treatment of the finite square well problem by L\&L, and has also kept abreast of the literature, deserves our apologies, for nothing is more tedious than reading an account of the errors one has not committed. Our motivation for writing this article has been to present an alternative graphical representation; to stress that the exercise of finding the eigenenergies of the finite well scrutinized by us will serve as an antidote to the tyro who may otherwise be tempted into thinking of {\sf Goal Seek} and {\sf Solver} as magic wands; and to suggest that those who teach (or use) iteration should obey the commandment ``Thou shalt not terminate (kill?) an iteration prematurely". 

\end{multicols}

%}     %ending \large

\begin{multicols}{2}

\end{multicols}
\section*{Appendix I: Iterative Solution of the Equations of Landau and Lifshitz }

Since all necessary information is readily available \cite{MurphyPhillips1976AJP}, we will only state that, for the problem of a well with $U = 25 \mbox{ eV}$ and $L = 0.5 \mbox{ nm}$, the five roots ($\xi_k$, $k=$1--5) of Eqs.~\ref{eq:Landau2} and ~\ref{eq:Landau3} were determined by transforming these relations into those listed below:

\begin{minipage}[b]{0.45\linewidth}
\begin{align}
\xi  &=f_1(\xi) \equiv \cos^{-1}(\gamma \xi), \tag{22:1}\\ 
\xi  &=f_3(\xi) \equiv \cos^{-1}(\gamma \xi) + \pi , \tag{22:3}\\
\xi  &=f_5(\xi)\equiv \cos(\xi)/\gamma \tag{22:5} ;
\end{align}
\end{minipage}
\hspace{0.5cm}
\begin{minipage}[b]{0.45\linewidth}
\begin{align}
\xi &=f_2(\xi) \equiv \sin^{-1}(\gamma \xi)+\pi, \tag{23:2}\\
\xi &=f_4(\xi) \equiv \sin^{-1}(\gamma \xi)+2\pi.\tag{23:4}
\end{align}
\end{minipage}

\vspace{2ex}

\noindent
The output of the Excel worksheet prepared by the authors is displayed in Figure~\ref{fig:LindbergExcel}.

\vspace*{2ex}

}% ending \large

\begin{figure}[h!] % float placement: (h)ere, page (t)op, page (b)ottom, other (p)age
  \centering
  % file name: F:/Sigmund FSW/LindbergExcel.png
  \includegraphics[width=16cm,height=10.3cm,keepaspectratio]{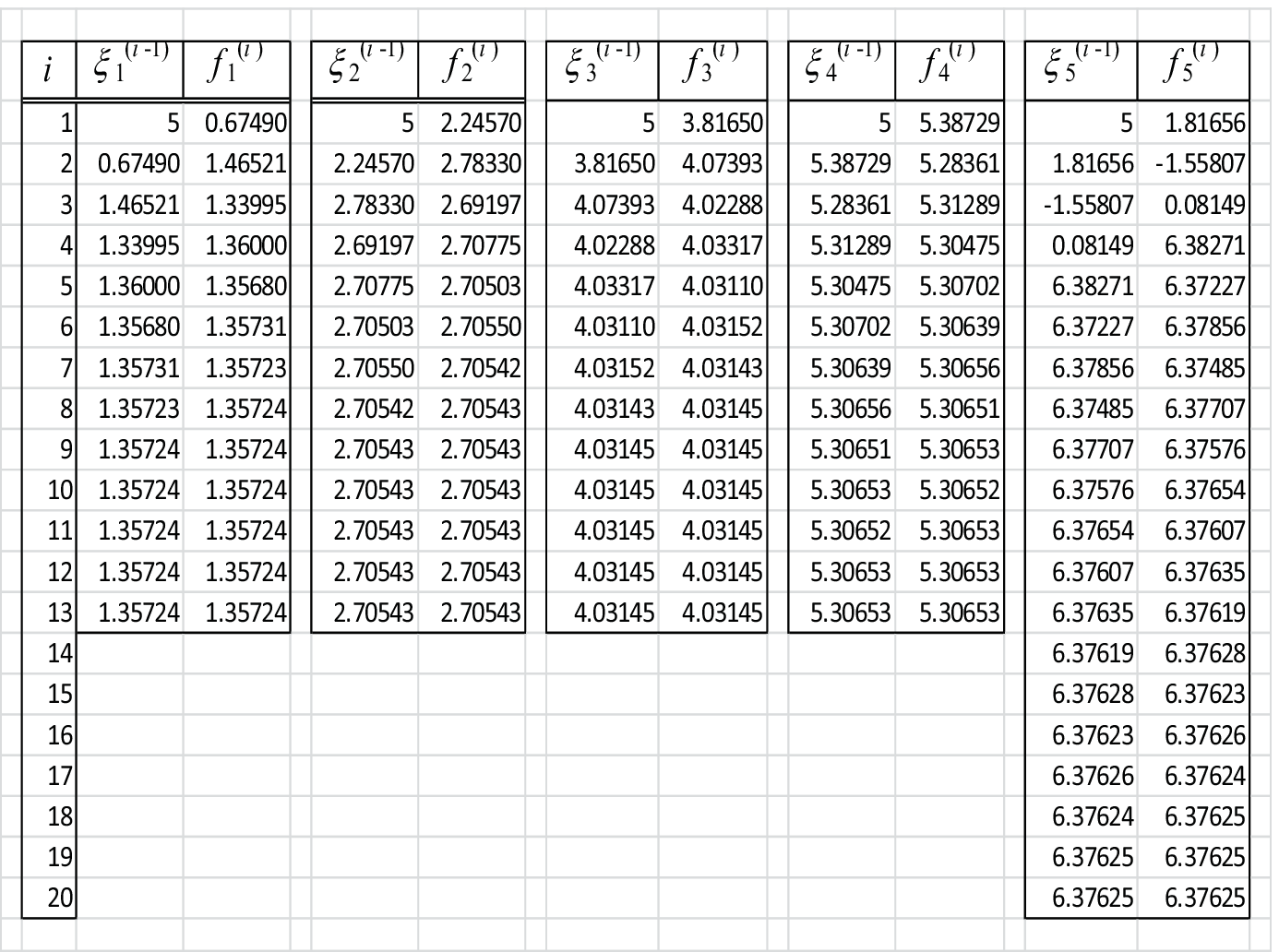}
  \caption{Excel output showing iterative calculations of the roots of Eqs.~\ref{eq:Landau2} and \ref{eq:Landau3}.}
  \label{fig:LindbergExcel}
\end{figure}

\newpage
\bibliographystyle{unsrt}
%\bibliography{FiniteWell}

\end{document}